# The Classical Harmonic Vibrations of the Atomic Centers of Mass with Micro Amplitudes and Low Frequencies Monitored by the Entanglement between the Two Two-level Atoms in a Single mode Cavity

Yong-Yi Huang†

*MOE Key Laboratory for Nonequilibrum Synthesis and Modulation of Condensed Matter,*
*Department of Optic Information Science and Technology,*
*Xi'an Jiaotong University, Xi'an 710049, China*

**Abstract**

We study the entanglement dynamics of the two two-level atoms coupling with a single-mode polarized cavity field after incorporating the atomic centers of mass classical harmonic vibrations with micro amplitudes and low frequencies. We propose a quantitative vibrant factor to modify the concurrence of the two atoms states. When the vibrant frequencies are very low, we obtain that: (i) the factor depends on the relative vibrant displacements and the initial phases rather than the absolute amplitudes, and reduces the concurrence to three orders of magnitude; (ii) the concurrence increases with the increase of the initial phases; (iii) the frequency of the harmonic vibration can be obtained by measuring the maximal value of the concurrence during a small time. These results indicate that even the extremely weak classical harmonic vibrations can be monitored by the entanglement of quantum states.

†Corresponding author, Email: yyhuang@mail.xjtu.edu.cn

## I. Introduction

The entanglement of quantum states discovered by Einstein Podolsky Rosen(EPR)[1] and Schrödinger[2] is one of the strangest phenomena in quantum mechanics. Bohm presents a vivid sample of quantum state entanglement, i.e. the entanglement of two electrons spin states[3]. Bell accepts EPR viewpoint and proposes Bell inequalities to give a judgment which theory describes the real world, quantum mechanics or local hidden variable theory[4]. Entanglement as a new resource can not only be applied to information field, such as quantum state teleportation[5], quantum cryptography[6], quantum dense coding[7], quantum computing[8] etc, but also be used to provide a new angle of view, such as the emergence of classicality[9], disordered systems[10], superconductivity[11] and superradiance[12] etc. The research about entanglement criteria has been widely carried out. Peres-Horodecki theorem[13,14] governs the entanglement of discrete states and continuous variables states, and bipartite system Gaussian states' criteria due to Duan[15] and Simon[16] is especially useful in laboratory. The concurrence of two qubits by Wootters[17] and the negativity [18,19] present quantitative descriptions of entanglement. Entanglement has been produced in laboratories, such as six or eight ions[20,21], six particles or ten qubits entanglement via photons[22,23] and nuclear and electron spins entanglement in diamond[24]. To meet the experimental needs, different methods for entanglement detection have been proposed[25, 26].

In this paper we study the effects of the centers of mass classical harmonic vibrations on the entanglement between the two two-level atoms in a single-mode polarized cavity field. Why do we select the classical harmonic vibrations with micro amplitudes and low frequencies? The first reason is that the vibration occurs widely in nature, any mass subject to a force in stable equilibrium acts as a harmonic oscillators. The second reason is that the detection problem of a gravitational wave is not solved yet. As we know, a gravitational wave is an extremely weak wave. A plane gravitational wave with two polarization states travelling in the z direction will deform the particles around a circle in the xy plane, a gravitational wave can be confirmed by measuring the oscillatory motions hit by a plane gravitational wave. We expect that the present results probably provide a new detection principle for a gravitational wave. The paper is organized as follows. In section **II** Wootters concurrence of the two different initial states are calculated. In section **III** we pay close attention to the vibrant factor for the centers of mass classical harmonic vibrations, which gives a modification of the concurrence of the two atoms states. In section **IV** a summary is presented.

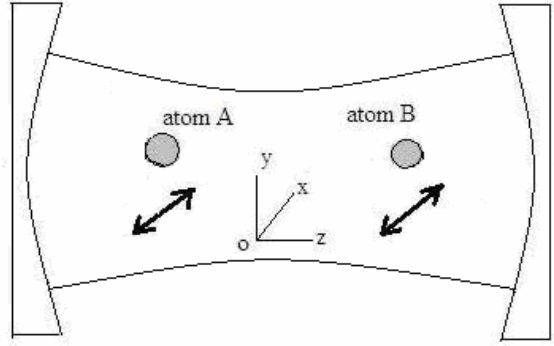

Fig.1 Schematic of a setup illustrates how the classical harmonic vibrations of centers of mass with micro amplitudes and low frequencies affect the concurrence between two two-level atoms coupling with a single-mode cavity field.

**II. The Calculations of Wootters Concurrence**

The system we study is shown in Fig.1. The Two same two-level A atom at $z_A = -z_0$ and B atom at $z_B = z_0$ are coupled to a single-mode cavity field polarized along y direction, which runs along z direction. The two atoms are controlled to harmonically vibrate along x direction by some drive. The system Hamiltonian is written as $H = H_0 + H_{CM} + H_I$, where $H_0$ describes the energy levels of the two atoms and the cavity, $H_{CM}$ describes vibrant centers of mass of the two atoms and $H_I$ describes the atom-field interaction. Under the rotating-wave approximation, $H_0, H_{CM}$ and $H_I$ are respectively given by

$$H_0 = \frac{1}{2}\hbar\omega_A \sigma_A^z + \frac{1}{2}\hbar\omega_B \sigma_B^z + \hbar\omega a^\dagger a,$$

$$H_{CM} = -\frac{\hbar^2}{2M_A}\frac{d^2}{dX_A^2} + \frac{1}{2}M_A\Omega_A^2 X_A^2 - \frac{\hbar^2}{2M_B}\frac{d^2}{dX_B^2} + \frac{1}{2}M_B\Omega_B^2 X_B^2$$

$$H_I = \hbar g \sum_{i=A,B} [a\sigma_i^+ \exp(ikz_i) + a^\dagger \sigma_i^- \exp(-ikz_i)].$$

In the total Hamiltonian, $a^\dagger, a$ are the bosonic operators; $\sigma^z$, $\sigma^+ = |e><g|$ and $\sigma^- = |g><e|$ are respectively Pauli operator, raising and lowering operators for the two-level atoms; $\Omega_A, \Omega_B$ are the vibrant frequencies of the two atoms A,B, $M_A, M_B$ are the masses of A, B atoms, $g$ is the coupling coefficient and $k$ is the wavenumber of the cavity field. In our thought experiment we assume that the vibrations of A and B atoms are fully the same. Without loss of generality we assume the conditions $\omega_A = \omega_B = \omega, M_A = M_B = M$, $\Omega_A = \Omega_B = \Omega$ are satisfied. The vibrations of the two atoms are supposed to have small amplitudes, the effect of the vibrations of the atomic centers of mass on the atom energy levels can be neglected. The recoil motions of A and B atoms along x direction are ignored when a photon is absorbed or emitted by A (or B) atoms. The reason is that A (or B) atom mass is extremely large compared with the absorbed (or emitted) photon energy. So we do not have to take the coupling between the vibrations of the atomic centers of mass and the single-mode polarized cavity field into account.

Several authors have studied the question that the two two-level atoms are coupled to a single-mode cavity field [27, 28, 29]. The two atoms, the single-mode polarized field and the two atoms form a closed system, the evolution equation of the system reads

$$\rho(t) = U(t)\rho(0)U^\dagger(t),$$

where the time evolution operator is $U(t) = \exp[-iHt/\hbar]$. Due to the relationships $[H_{CM}, H_0] = [H_{CM}, H_I] = 0, [H_{CM} + H_0, H_I] = 0$, we have

$$U(t) = \exp[-iH_I t/\hbar]\exp[-iH_0 t/\hbar]\exp[-iH_{CM} t/\hbar] \tag{1}$$

The reduced density for the two atoms is given by

$$\rho(t)_{atoms} = Tr_E[U(t)\rho(0)U^\dagger(t)]. \tag{2}$$

The trace $Tr_E$ in Equ (2) includes the traces over both cavity field and vibrations of the atomic centers of mass. Taken Equ(1) into account, $\rho(t)_{atoms}$ in the interaction picture is written as

$$\rho(t)_{atoms} = Tr_E[e^{-iH_I t/\hbar}\rho(0)e^{iH_I t/\hbar}]. \tag{3}$$

We still use the previous symbols throughout this paper not to cause confusion. $e^{-iH_I t/\hbar}$ is exactly worked out in the atomic basis $\{|ee>, |eg>, |ge>, |gg>\}$ similarly in ref.[28], where $|e>$ is excited state and $|g>$ is ground state, i.e.

$$e^{-iH_I t/\hbar} = \begin{pmatrix} 2g^2 a(C-\Theta)a^+ + 1 & -igaSe^{ikz_0} & -igaSe^{-ikz_0} & 2g^2 a(C-\Theta)a \\ -igSa^+ e^{-ikz_0} & (\cos\Gamma t + 1)/2 & (\cos\Gamma t - 1)e^{-2ikz_0}/2 & -igSae^{-ikz_0} \\ -igSa^+ e^{ikz_0} & (\cos\Gamma t - 1)e^{2ikz_0}/2 & (\cos\Gamma t + 1)/2 & -igSae^{ikz_0} \\ 2g^2 a^+(C-\Theta)a^+ & -iga^+ Se^{ikz_0} & -iga^+ Se^{-ikz_0} & 2g^2 a^+(C-\Theta)a + 1 \end{pmatrix} \quad (4)$$

Here $\Gamma^2 = \Theta^{-1} = 2g^2(2a^+ a + 1)$ and C and S are defined by $C = \Theta\cos\Gamma t$ and $S = \Gamma^{-1}\sin\Gamma t$.

We study a typical initial states and give the correction of the vibrations of the atomic centers of mass on the concurrence between the two atoms states[17]. The initial state is

$$\rho(0) = |gg><gg| \otimes |1><1| \otimes |\psi_n \psi_m><\psi_n \psi_m|, \quad (5)$$

where $|1>$ and $|\psi_n \psi_m>$ denote the one photon state and the eigenstates for the two atoms. Substituting $\rho(0)$ from Equ (5) into Equ (3) and inserting the completeness $\int |X_1><X_1| \otimes |X_2><X_2| dX_1 dX_2 = I$ into the traces over the vibrations, we obtain

$$\rho(t)_{atoms} = \begin{pmatrix} 0 & 0 & 0 & 0 \\ 0 & \dfrac{\sin^2 \sqrt{2}gt}{2} & \dfrac{\sin^2 \sqrt{2}gt}{2} & 0 \\ 0 & \dfrac{\sin^2 \sqrt{2}gt}{2} & \dfrac{\sin^2 \sqrt{2}gt}{2} & 0 \\ 0 & 0 & 0 & \cos^2 \sqrt{2}gt \end{pmatrix} \quad (6)$$

$$\times \int <\psi_n \psi_m | X_1 X_2><X_1 X_2|\psi_n \psi_m> dX_1 dX_2$$

Quantum number $n,m$ will be very large under the low frequencies condition of $\Omega << g, \Omega << \omega$, the probability $\int <\psi_n \psi_m | X_1 X_2><X_1 X_2|\psi_n \psi_m> dX_1 dX_2$ can be regarded as classical harmonic oscillators probabilities. The condition $\Omega << g, \Omega << \omega$ can be satisfied in laboratory. In fact the typical frequency $\Omega$ of mechanical vibration due to a gravitational wave is about $10^3$ Hz[30], a strong coupling coefficient $g$ can arrive at $10^6$ Hz[31], and the resonant $\omega$ is higher than the coupling coefficient $g$. In order to calculate the classical harmonic oscillators' probabilities, we use the equations of the vibrations. Given $\alpha = \sqrt{M\Omega/\hbar}$ and $\xi = \alpha x$, we obtain the classical motion equations of the two atoms: $\xi_1 = A_1 \sin(\Omega t + \delta_1)$ and $\xi_2 = A_2 \sin(\Omega t + \delta_2)$, where $A_1$ and $A_2$ are the classical amplitudes, $\delta_1, \delta_2$ are the initial phases of the two atoms. The classical harmonic oscillator probability density is $w(\xi) = <\psi|\xi><\xi|\psi> = \frac{1}{\pi\sqrt{A^2-\xi^2}}$, and $w(\xi)$ increases with the increase of the displacement $\xi$ in $\xi \in [0, A]$. We do not consider $\xi \in [-A, 0]$ because of the classical harmonic oscillator probability density's symmetry between $[0, A]$ and $[-A, 0]$.

We work out $\int <\psi_n\psi_m|\xi_1\xi_2><\xi_1\xi_2|\psi_n\psi_m> d\xi_1 d\xi_2$ during a very short time $\Delta t$ i.e.

$$\int_{\xi_{10}}^{\xi_{10}+\zeta_1} w(\xi_1)d\xi_1 \int_{\xi_{20}}^{\xi_{20}+\zeta_2} w(\xi_2)d\xi_2$$
$$= \frac{1}{\pi^2}[\arcsin(\frac{\zeta_1}{A_1}+\sin\delta_1)-\delta_1][\arcsin(\frac{\zeta_2}{A_2}+\sin\delta_2)-\delta_2] \quad (7)$$

where $\zeta_1, \zeta_2$ denote the absolute displacements of the two atoms, and $\delta_1$, $\delta_2$ are the initial phases, that is, $\xi_{10} = A_1\sin\delta_1, \xi_{20} = A_2\sin\delta_2$ with $\xi_{10}, \xi_{20}$ denoting the initial displacements. $\Delta t$ is satisfied the condition $1/g << \Delta t << 1/\Omega$. The condition guarantees that the integral upper limit and lower limit in Equ(7) are $[\xi_{10}, \xi_{10}+\zeta_1], [\xi_{20}, \xi_{20}+\zeta_2]$ rather than $(-\infty, +\infty)_1, (-\infty, +\infty)_2$. During a very short time $\Delta t$ the relative displacements are very small and satisfy the relationships $0 < \zeta_1/A_1 << 1, 0 < \zeta_2/A_2 << 1$.

From Equ. (6) and Equ. (7), we can obtain the concurrence $C(\rho) = \max(0, \sqrt{\lambda_1} - \sqrt{\lambda_2} - \sqrt{\lambda_3} - \sqrt{\lambda_4})$, where the quantities $\lambda_i$ are the eigenvalues of the matrix $\rho(\sigma_A^y \otimes \sigma_B^y)\rho^*(\sigma_A^y \otimes \sigma_B^y)$ arranged in decreasing order. $\rho^*$ is the complex conjugation of $\rho$ in the atomic basis $\{|ee>, |eg>, |ge>, |gg>\}$, and $\sigma_A^y \otimes \sigma_B^y$ is the direct product of Pauli matrix expressed in the same basis[32]. The concurrence is calculated as

$$C(\rho) = \frac{1}{\pi^2}[\arcsin(\frac{\zeta_1}{A_1}+\sin\delta_1)-\delta_1][\arcsin(\frac{\zeta_2}{A_2}+\sin\delta_2)-\delta_2]\sin^2\sqrt{2}gt \quad (8)$$

### III. The Vibrant Factor for the Classical Harmonic Vibrations of the Atomic Centers of Mass

Wootters concurrences Equ(8) has a vibrant factor for the classical harmonic vibrations of the atomic centers of mass

$$K(\delta_1, \delta_2) = \frac{1}{\pi^2}[\arcsin(\frac{\zeta_1}{A_1}+\sin\delta_1)-\delta_1][\arcsin(\frac{\zeta_2}{A_2}+\sin\delta_2)-\delta_2] \quad (9)$$

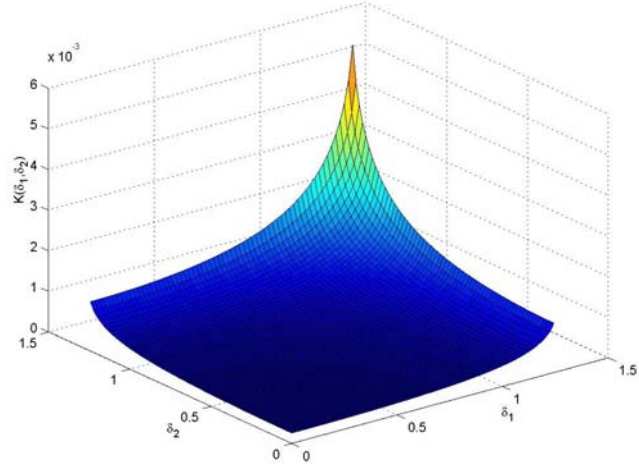

Fig. 2 The vibrant factor $K$ is versus the initial phases $\delta_1$, $\delta_2$, with $\zeta_1/A_1 = \zeta_2/A_2 = 0.05$.

The factor $K$ versus the initial phases $\delta_1$, $\delta_2$ is shown in Fig. 2, which is our main result. Seen from Fig.2 and Equ(9) we obtain three results: (1) The vibrant factor will increase with the increase of the initial phases $\delta_1, \delta_2$. The reason is that the initial phases $\delta_1, \delta_2$ correspond to the different displacements $\xi_1, \xi_2$, the probability density $w(\xi)$ increases with the increase of displacement $\xi$. (2) The vibrant factor for the vibrations of the atomic centers of mass depends on the relative vibrant displacements $\frac{\zeta_1}{A_1}, \frac{\zeta_2}{A_2}$ and the initial phases $\delta_1$, $\delta_2$, rather than the absolute vibrant amplitudes $A_{1,2}$. (3) The harmonic vibrations of atoms centers of mass greatly reduce the concurrence to three orders of magnitude. Because in the condition of $\Omega \ll g, \Omega \ll \omega$ the harmonic oscillator probabilities are not normalized during a very short time $\Delta t$, the probabilities within the small relative displacements $\frac{\zeta_1}{A_1}, \frac{\zeta_2}{A_2}$ are of course much smaller than unity, please see Equ (7). Actually the relative displacements $\frac{\zeta_1}{A_1}, \frac{\zeta_2}{A_2}$ depends on arbitrary short time interval $\Delta t$, which satisfies the condition $1/g \ll \Delta t \ll 1/\Omega$. Given $\frac{\zeta_1}{A_1} = \frac{\zeta_2}{A_2} = 0.05$ and $g = 10^6 Hz$, $\Omega = 10^3 Hz$, we have $\sin\delta_1 = \sin\delta_2 = 0.95$, i.e. $\delta_1 = \delta_2 \simeq 1.25$ maximally due to the fact $\frac{\zeta}{A} + \sin\delta = 1$. Substituting $\delta_1 = \delta_2 \simeq 1.25$ into $K(\delta_1, \delta_2)$, we obtain that $K(\delta_1, \delta_2)$ maximum is about 0.01.

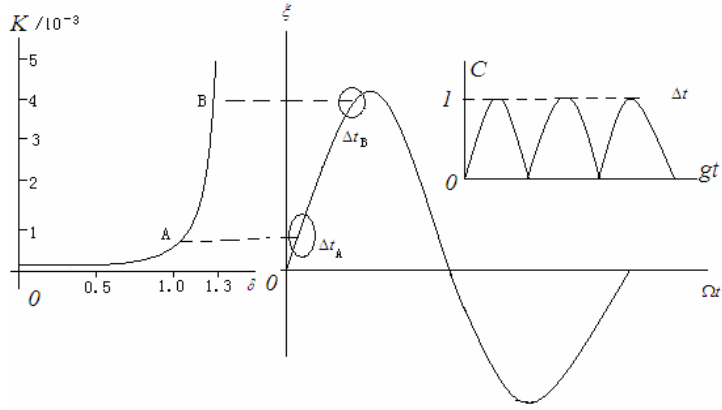

Fig. 3 It is shown the process of the classical harmonic vibrations of the atomic centers of mass monitored by the entanglement between the two two-level atoms in a single mode cavity. The vibrant factor $K$ versus the two atomic initial phases $\delta_1, \delta_2$ lies in the left panel, with $\delta_1 = \delta_2$ and $\zeta_1/A_1 = \zeta_2/A_2 = 0.05$ without loss of generality. The harmonic amplitude with time in the right panel corresponds to the left vibrant factor. The inset in the right panel is the concurrence $C$ versus $gt$ without considering the harmonic vibrations of the atomic centers of mass. A dot in the left panel corresponds to a measurement time $\Delta t$ with an initial phase $\delta$, and it takes a different time $\Delta t$ at a different phase $\delta$ to keep the relative displacement $\zeta/A$ to be constant. For instance, dot A and dot B in the left panel have different initial phases and correspond to different displacements in the right panel. If the relative displacement $\zeta/A$ is kept to be constant, then the measurement time in dot A is less than the time in dot B i.e. $\Delta t_A < \Delta t_B$. The vibrant factor $K(A)$ is less than $K(B)$ because of the probability density in dot A less than the probability density in dot B seen from Equ.(7).

Now we discuss how to monitor the classical harmonic vibrations of the atomic centers of mass by measuring the entanglement between the two two-level atoms in a single mode cavity. Without considering the classical harmonic vibrations of the atomic centers of mass, the concurrence versus time between the two two-level atoms in a single mode cavity is very simple, i.e. $C(\rho)_0 = \sin^2 \sqrt{2} gt$, shown in the inset of the right panel in Fig 3. After incorporating the vibrations of the atomic centers of mass, the concurrence has a vibrant factor $K$ in $C(\rho)_0$, i.e. Equ(8). The selection of a measurement time $\Delta t$ is very subtle, $1/g \ll \Delta t \ll 1/\Omega$ is required. It indicates that $\Delta t$ is much less than the period $2\pi/\Omega$ of the classical harmonic vibrations, however, much larger than the period $\pi/(\sqrt{2}g)$ of the $C(\rho)_0$. Because the maximal value of $C(\rho)_0$ is unity, theoretically we can obtain the vibrant factor $K$ by measuring the maximal value of the concurrence $C(\rho)$ during a time $\Delta t$. In one period $2\pi/\Omega$ of the classical harmonic vibrations, once we obtain the vibrant factor $K$ versus the initial phases $\delta_1, \delta_2$ of the two atomic centers of mass, i.e. the left panel $K$ in Fig 3, we confirm the existence of the classical harmonic vibrations of the two atoms. It is very valuable that the vibrant factor $K$ is independent of the absolute amplitudes, which indicates we can confirm the existence of the classical harmonic vibrations with micro amplitudes. If we keep each measurement time $\Delta t$ to be constant, we obtain the vibrant factor $K$ as

$$K(\delta_1, \delta_2) = 4\frac{\Delta t^2}{T^2}, \tag{10}$$

where $T$ is the frequency of the classical harmonic vibration. Equ (10) is even independent of the initial phases. In practice we should divide Equ(10) by 4,

$$K(\delta_1, \delta_2) = \frac{\Delta t^2}{T^2}, \tag{11}$$

because each relative vibrant displacement $\zeta_{1,2}/A_{1,2}$ contains two to-and-fro processes in one-half period shown in Fig 3. We can obtain the vibrant factor $K$ by measuring the maximal value of the concurrence $C(\rho)$ during a time $\Delta t$, so we acquire the frequency of a classical harmonic vibration even with an amplitude tending to zero.

**IV. Summary**

In conclusion we have studied the entanglement of the two two-level atoms coupled with a single-mode polarized cavity field after incorporating the classical harmonic vibrations of the atomic centers of mass. Our results bring an interesting byproduct, readers always mention that there are not the effects of the motions of the atomic centers of atoms on the entanglement of the two atoms because the vibrations of the atomic centers of mass fully decouple with the internal freedoms of atoms. Here we give some reasons why the argument is not always right. In some conditions the decoupled motions of atoms with mode cavity field and atoms internal freedoms indeed have the effects atoms on the entanglement of atoms internal freedoms. The conditions are that the frequencies of atoms classical harmonic vibrations are very low i.e. their period is rather large, and that experimenters measure the entanglement of atoms internal freedoms during a very short time compared with one period. Of course, if experimenters measure the entanglement during a whole period of atoms harmonic vibrations, the decoupled motions of atoms do not have anything to do with the entanglement of the internal freedoms.

Here we give a familiar example: A plant may produce new flowers, man is young but once. Man has a longer period than a plant, that is to say, a man eighty years old should observe that a plant produces new flowers eighty times. In the paper the time-dependent entanglement of atoms internal freedoms is just the plant, the harmonic vibrations of atoms are the man. There are many periods of the entanglement in one period of atoms harmonic vibrations. Experimenters will observe many times varieties of the entanglement even during a small time of one period of the atoms vibrations. In this paper, a vibrant factor has been derived to quantitatively describe the effects of the low-frequency vibrations of the atoms centers of mass on the entanglement of the internal freedoms of atoms.

The classical harmonic vibrations of centers of mass reduce the concurrence to three orders of magnitude. The concurrence is sensitively affected by the initial phases and relative displacements rather than the absolute vibrant amplitudes under the condition $\Omega << g, \Omega << \omega$.

The larger the initial phases become, the larger the concurrence becomes. Measuring phase-varying entanglement can confirm the existence of micro vibrations. If we keep each measurement time $\Delta t$ to be constant, we even obtain the frequency of the harmonic vibrations by measuring the maximal value of the concurrence during a small time $\Delta t$. The vibrant factor *K* is independent of the absolute amplitudes, our results maybe provide a new principle for the detection of gravitational wave.